\documentclass[11pt]{JHEP3}

\usepackage{cite}
\usepackage{epsf}
\usepackage{latexsym}
\usepackage{epsfig,multicol}

\def\preal{{\rm Re\,}}
\def\pim{{\rm Im\,}}

\def\yzero{\smash{\hbox{$y\kern-4pt\raise1pt\hbox{${}^\circ$}$}}}
\def\p{\partial}
\def\a{\alpha}
\def\b{\beta}
\def\g{\gamma}
\def\d{\delta}
\def\be{\begin{equation}}
\def\ee{\end{equation}}
\def\bea{\begin{eqnarray}}
\def\eea{\end{eqnarray}}
\def\Om{\Omega}

\def\th{\theta}

\def\-{\hphantom{-}}
\def\ov{\overline}
\def\s2{\frac{1}{\sqrt2}}

\def\oh{\frac{1}{2}}
\def\be{\begin{equation}}
\def\ee{\end{equation}}
\def\bea{\begin{eqnarray}}
\def\eea{\end{eqnarray}}

\def\IF{\relax{\rm I\kern-.18em F}}
\def\II{\relax{\rm I\kern-.18em I}}
\def\IP{\relax{\rm I\kern-.18em P}}
\def\IC{\relax\hbox{\kern.25em$\inbar\kern-.3em{\rm C}$}}
\def\IR{\relax{\rm I\kern-.18em R}}

\def\cn{{\cal N}}
\def\cm{{\cal M}}

\def\cf{{\cal F}}

\def\Dsl{\,\raise.15ex\hbox{/}\mkern-13.5mu D} 
\def\IZ{Z\kern-.4em  Z}

\newcommand{\eqn}[1]{(\ref{#1})}
\newcommand{\mt}[1]{\textrm{\tiny #1}}
\newcommand{\mbar}{{\ov{m}}}
\newcommand{\sac}{\, , \qquad}
\newcommand{\cy}{CY$_3$ }

\def\T{{\bf T}}

\def\car{{\cal R}}

\def\cl{{\cal L}}
\def\cs{{\cal S}}

\def\Lam{\Lambda}
\def\D{\Delta}
\def\G{\Gamma}
\def\raw{\rightarrow}

\def\k{\kappa}

\def\Sig{\Sigma}

\def\cw{{\cal W}}
\def\z{\zeta}

\newcommand{\ZZ}{\mathbb{Z}}
\newcommand{\CC}{\mathbb{C}}

\newcommand{\RR}{\mathbb{R}}
\newcommand{\NN}{\mathbb{N}}

\title{An Open String Landscape}

\author{Jaume Gomis,$^{1}$ Fernando Marchesano,$^{2,3}$ and David Mateos$^{1}$\\
      $^{1}$\ Perimeter Institute for Theoretical Physics, Waterloo, Ontario N2L 2Y5, Canada\\
      $^{2}$\ Department of Physics, University of Wisconsin-Madison, WI 53706, USA \footnote{Permanent address} \\
      $^{3}$\ Fields Institute,  Toronto, Ontario, M5T 3J1, Canada \\ \\
        E-mail: \email{jgomis@perimeterinstitute.ca}, \email{marchesa@physics.wisc.edu}, \email{dmateos@perimeterinstitute.ca}
}

\received{\today}               
\revised{}
\accepted{}               

\preprint{hep-th\0506179}
\preprint{MAD-TH-05-4}

\abstract{The effect of fluxes on open string moduli is studied by analyzing the constraints imposed by supersymmetry on D-branes in type IIB flux backgrounds. We show that generically the conditions of supersymmetry cannot be maintained when moving along the geometrical moduli space of the brane, so that open string moduli are lifted. We argue that there is a disconnected and discrete set of supersymmetric solutions to the open string equations of motion, which extends the familiar closed string landscape to the open string sector.}

\begin{document}

\section{Introduction and Conclusions}

One of the lessons of the second superstring revolution is that
there is a unique underlying theory, namely M-theory. Recent work
on flux vacua \cite{gvw,drs,gkp,kst}, however, supports
the view that the theory has a very large \cite{bp,counting}
number of ground states,\footnote{At least in the supergravity
approximation in which they are found.} generically dubbed as the
landscape of string vacua \cite{landscape}. Thus far, most of the
analysis has focused on studying, for a given background geometry,
the critical points of the potential as a function of the closed
string moduli and flux quanta. It has been found that,
generically, the complex structure closed string moduli are fixed
by the flux-induced perturbative superpotential, while the K\"ahler
moduli have flat directions which nevertheless may be lifted by
non-perturbative effects.

In this paper we analyze the analogous problem for open string
moduli in type IIB flux compactifications. We derive the
conditions for D-branes to preserve supersymmetry, closely following
\cite{mmms}. Supersymmetry imposes constraints both on the
geometrical properties of the cycle wrapped by the D-brane, and on
the gauge invariant field strength ${\cal F}$ living on its
worldvolume. In the absence of background fluxes, the open string
moduli arise from the space of geometrical deformations that
preserve the holomorphicity of the cycle, as well as from Wilson
lines. In the presence of fluxes, we show that the geometrical
moduli are generically lifted, while the Wilson lines remain flat
directions. This parallels the closed string analysis, in which
the perturbative superpotential generically fixes the complex
structure moduli, but not the K\"ahler moduli.

From the viewpoint of the four-dimensional effective field theory
description, the conditions imposed by supersymmetry on the open
string moduli should be interpreted as arising from a flux-induced
superpotential for the open string fields. We find that, 
generically, there is a
discretum of supersymmetric open string vacua disconnected from
each other, since D-branes can be placed in a supersymmetric way
only at a discrete set of points. This increases the number of
choices that must be made in order to specify a vacuum, but now
the choices are in the open string sector. We have named this
space of choices the open string landscape. It would be
interesting to map out this sector in more detail and try to
extract features that could be of phenomenological interest.

Our analysis may also shed light on the problem of closed string
moduli stabilization. Supersymmetric D-brane instantons can
generate a non-perturbative superpotential that stabilizes the
volume modulus of the compactification manifold, which is left
unfixed by the flux-induced perturbative part. Generation of a
superpotential by one such instanton depends on the precise number
of fermionic zero-modes on its worldvolume. Supersymmetry implies
that a reduction in the number of bosonic zero-modes in the
presence of fluxes must be accompanied by the corresponding
reduction in the number of their fermionic counterparts.
Therefore, an instanton that does not contribute in the absence of fluxes
may now contribute to the superpotential.

The plan of the paper is as follows. In Section \ref{kappaD} we
employ the $\kappa$-symmetric formulation of D-branes to derive
the conditions imposed on them by unbroken supersymmetry in the
presence of fluxes. We express these conditions in terms of the
geometry of the cycle wrapped by the brane and its worldvolume
flux ${\cal F}$. In section \ref{local} we show that, generically,
the supersymmetry condition on ${\cal F}$ reduces the would-be
moduli space associated to deformations of the cycle to a set of
isolated points. We also propose an interpretation of the
supersymmetry conditions in terms of a flux-generated
superpotential for the open string moduli. In Section \ref{global}
we illustrate the general analysis of section
\ref{local} with a simple model, in which the appearance of
an open string landscape can be seen explicitly. We discuss
some generic features of this new landscape of vacua.
Section \ref{app} contains some formal and phenomenological applications
of our analysis. Some calculations are relegated to the Appendix.

\section{Supersymmetric D-branes in Flux Backgrounds}
\label{kappaD}

In this section, we determine the brane embeddings that preserve supersymmetry by doing a D-brane worldvolume analysis. In this approach, the supersymmetries preserved by a D-brane in a $\cn \geq 1$ supergravity background are those generated by background Killing spinors $\eta$ that satisfy the condition \cite{BKOP}
\be
\Gamma_{\kappa} \eta = \eta \,,
\label{kappa}
\ee
where $\Gamma_{\kappa}$ is the traceless, unit-square matrix that appears in the $\kappa$-symmetry transformation of the D-brane worldvolume fermions \cite{action}. This approach has the benefit that it can be applied to any supergravity background, in particular to supersymmetric backgrounds with fluxes.

In this paper we will concentrate on type IIB D$p$-branes, for which $\Gamma_{\kappa}$ takes the form\footnote{In our conventions, the 32 supersymmetry generators of type IIB supergravity are grouped into a doublet of real Majorana-Weyl
 fermions. The Pauli matrices act on this doublet index.}
\be
\Gamma_{\kappa} = \frac{\sqrt{|g|}}{\sqrt{|g+\cf|}} \sum_{n=0}^\infty
\frac{1}{2^n n!} \gamma^{\mu_1 \nu_1 \ldots \mu_n \nu_n}
\cf_{\mu_1 \nu_1} \ldots \cf_{\mu_n \nu_n} (-1)^n (\sigma_3)^{n+\frac{p-3}{2}}
\sigma_2 \, \otimes \Gamma_{(0)} \,,
\label{kappamatrix}
\ee
where\footnote{We work in Euclidean signature on the worldvolume and on target space. This allows us to study both instantonic and space-filling branes. For the case of Lorentzian space-filling D-branes, one can trivially analytically continue back to Lorentzian signature.}
\be
\Gamma_{(0)} = \frac{1}{(p+1)! \sqrt{g}} \,
\epsilon^{\mu_1 \ldots \mu_{(p+1)}}
\gamma_{\mu_1 \ldots \mu_{(p+1)}} \,.
\ee
In these formulas, $\mu, \nu=0,1,\ldots,p$ are worldvolume
indices, $g$ is the induced metric on the D-brane, and
\be
\cf = 2\pi \alpha' F + B
\label{f}
\ee
is the gauge-invariant field strength on the D-brane. Throughout this paper, a pull-back of spacetime forms onto the D-brane worldvolume is understood where necessary, such as that of the $B$-field in the equation above. The $\gamma$-matrices are induced from the spacetime $\Gamma$-matrices as
\be
\gamma_\mu = \partial_\mu Z^M \Gamma_M \,,
\ee
where $Z^M$ are local coordinates in spacetime.

Our present goal is to solve for the supersymmetric D-brane embeddings in warped type IIB flux backgrounds. It will be convenient to first present the analysis performed by Mari\~ no, Minasian, Moore, and Strominger (MMMS) in \cite{mmms}, where they analyze the embedding condition of D-branes in Calabi-Yau compactifications and in the presence of a closed $B$-field. As we will show, the conditions found in \cite{mmms} can be easily generalized to the case of our interest after some simple modifications.

\subsection{D-branes in a Calabi-Yau space with a closed $B$-field }

In \cite{mmms}, a general analysis of the conditions implied by eq.\eqn{kappa} was performed for type II D-branes embedded in special holonomy manifolds and with a closed $B$-field. In particular, it was analyzed the case of $\cn=2$ type IIB compactifications on Calabi-Yau three-folds (CY$_3$), for which the spacetime metric takes the form
\be
ds^2 = \d_{ab} \, dX^a dX^b
+ 2 G^\mt{CY}_{n\ov{m}} \, dZ^n dZ^{\ov{m}} \,,
\label{CY}
\ee
where $Z^m$ and $Z^{\ov{m}}$ are holomorphic and anti-holomorphic coordinates on CY$_3$. All the type IIB  field strengths vanish in this background but the role of the closed $B$-field cannot be ignored since it plays an important role on D-branes, as can be seen from equation \eqn{kappamatrix}.

The first step is to write down the background Killing spinor appearing in \eqn{kappa}. In this background, the associated Killing spinors $\eta$ are linear combinations of the two covariantly constant spinors of the CY$_3$, $\eta_+$ and $\eta_-$. These are complex conjugate to one another, are normalized such that $\eta_\pm^\dagger \eta_\pm =1$, and satisfy the usual relation\footnote{Unless stated explicitly, all $\Gamma$-matrices we write down are curved space ones.}
\be
\Gamma_{\mbar} \, \eta_+ = 0 \sac \Gamma_m \, \eta_-=0 \,.
\label{annihilation}
\ee
The K\"ahler two-form and the holomorphic three-form of the \cy can be defined in terms of these spinors through the relations
\be
\Gamma_{m\bar{n}} \, \eta_+ = i J_{m\bar{n}} \, \eta_+ \sac
\Gamma_{mnp} \, \eta_+ = \Omega_{mnp} \, \eta_- \,,
\label{jomega}
\ee
so that $dJ = d\Omega=0$.

Using the equations above, it was shown in \cite{mmms} that an Euclidean D$(2n-1)$-brane wrapped on a $2n$ cycle $\cs_{2n}$ of \cy preserves supersymmetry if and only if certain equations in terms of $\Om$ and $J$ are satisfied. These equations can be written as
\bea
&& \left. e^{iJ - \cf} \right|_{2n} = e^{i\th} \,
\frac{ \sqrt{ |g+\cf| }}{ \sqrt{|g| }} \, d {\rm vol}_{2n} \,,
\label{mmms1}\\
&& \left.\iota_m \Omega \wedge  e^{iJ - \cf}\right|_{2n} = 0 \quad m=1,2,3\,,
\label{mmms2}
\eea
where the subscript `$2n$' means that we keep only the form of degree $2n$ and $\iota_m$ denotes the interior contraction with $\partial/\partial Z^m$. Pull-backs of spacetime forms onto the worldvolume are again understood. Finally, $e^{i\th}$ is a phase that parametrizes the embedding of a $U(1)$ family of $\cn=1$ algebras inside the bulk $\cn=2$ superalgebra. The same result also applies to a D($2n+3$)-brane that fills the four flat directions in \eqn{CY} and wraps a $2n$ cycle in CY$_3$.

Equation (\ref{mmms2}) can be shown to be equivalent to the condition that \cite{mmms}
\be
\cs_{2n}\ {\rm is\ holomorphic} \sac  \cf^{(2,0)} = 0 \,.
\label{mmms2*}
\ee

\subsection{D-branes in a warped, flux background}

We would like now to consider D-branes in a class of supersymmetric type IIB backgrounds with the following spacetime metric
\be
ds^2 = \D (Z)^{-1}\, \d_{ab} \, dX^a dX^b
+ 2 G_{n\ov{m}} \, dZ^n dZ^{\ov{m}} \,.
\label{warped0}
\ee
The background is now warped by $\D (Z)$, and the compactification manifold $\cm_6$ does no longer posses $SU(3)$ holonomy, but belongs to the more general class of $SU(3)$-structure manifolds (see, e.g., \cite{glmd,ccdlmz,gmpt}). In addition, one may also consider background fluxes. As shown in \cite{gmpt}, in any supersymmetric compactification on a $SU(3)$-structure manifold one can define an Hermitian metric, a normalizable supersymmetry generator $\eta_{+}$, its complex conjugate $\eta_-$, and a couple of differential forms $J$ and $\Om$ such that the relations (\ref{jomega}) still hold. One can then check that the computations carried out in \cite{mmms} also apply to this case, and that the supersymmetry equations (\ref{mmms1}) and (\ref{mmms2}) still apply to this more general type IIB backgrounds with fluxes and/or torsion.\footnote{In $\cn=1$ supersymmetric backgrounds, the phase $\th$ in (\ref{mmms1}) is fixed by  the Killing spinor of the background.}

The simplest example of these class of vacua is obtained by taking $\cm_6$ to be a conformal Calabi-Yau, more precisely taking $G_{n\ov{m}} = \D(Z) G_{n\ov{m}}^{\mt{CY}}$. Supersymmetric solutions to this ansatz can be found when one allows a non-trivial self-dual five form flux $F_5$, as well as a $(2,1)$ and primitive  imaginary-self-dual (ISD) three form flux $G_3$ on the internal space. In addition sources of negative tension must be included when $\cm_6$ is compact. The presence of fluxes, both from the NS-NS and RR sector, can be taken into account in the computation of the supersymmetric D-brane embeddings \eqn{kappa}. The RR fluxes do not appear  directly in the form of the $\kappa$-symmetry projector \eqn{kappamatrix}. Nevertheless they enter in an important way in determing the Killing spinor of the corresponding background and thus affect the solution of \eqn{kappa}. The NS-NS fluxes enter both indirectly in the form of the background Killing spinor and directly in the form of the projector \eqn{kappamatrix}.

The Killing spinor of the warped Calabi-Yau background with ISD fluxes has the same structure as the one considered in \cite{mmms} modulo the extra
constraint:\footnote{Here $\Gamma_{0123}$ refers to flat
$\Gamma$-matrices. Recall that we are working in euclidean
signature, thus $\Gamma^2_{0123}=1$\label{euclid}.}
\be
\sigma_2\otimes\Gamma_{0123} \eta=\eta,
\label{killing}
\ee
which reflects the fact that we only have ${\cal N}=1$
supersymmetry. We note that the projection on the spinor is
precisely the one produced by a D3-brane. Just like in the
previous section, we can construct normalizable spinors
$\eta_\pm$ on the internal manifold, as well as the $\G$-matrices
 \be
 \eta_\pm = \D^{-1/4} \, \eta_{\pm}^\mt{CY} \sac
 \Gamma_a = \D^{-1/2} \, \Gamma_a^\mt{CY} \sac
 \Gamma_m = \D^{1/2} \, \Gamma_m^\mt{CY} \,,
\ee
where the superscript `CY' corresponds to the quantities in the
underlying Calabi-Yau geometry. It is now straightforward to
verify that the relations \eqn{annihilation} and \eqn{jomega}
still hold, with $J$ and $\Omega$ related to the corresponding
forms in the underlying CY geometry by
\be
 J = \D \, J^\mt{CY} \sac \Omega = \D^{3/2} \, \Omega^\mt{CY} \,.
\ee
Notice that $J$ and $ \Omega$ are no longer closed due to the warp factor dependence. It then follows that
the conditions for an Euclidean D-brane to preserve supersymmetry
in a conformal Calabi-Yau background take the form
\eqn{mmms1} and \eqn{mmms2}. The phase $\th$, however, is now fixed to $e^{i \th}=-1$ by the form of the Killing spinor (\ref{killing}).

The same conditions hold for a D$(2n+3)$-brane wrapped on
$\cs_{2n}$. This is because for such a D-brane the $\kappa$-symmetry matrix
factorizes as
 \be
 \Gamma = \Gamma_{\cs_{2n}} \, \Gamma_{0123} \otimes \sigma_2 \,,
\ee
where $\Gamma_{\cs_{2n}}$ is the matrix associated to an Euclidean
D-brane wrapped on $\cs_{2n}$. Since the Killing spinor satisfies
\eqn{killing} the conditions for the space-filling branes (filling Minkowski space)
take the same form as for the brane just wrapping $\cs_{2n}$. In
particular, we see that the supersymmetry conditions for a
space-filling D7-brane and an Euclidean D3-brane wrapped on the
same cycle $\cs_4$ are identical.

To summarize, the conditions for a D-brane to preserve
supersymmetry in a $SU(3)$-structure compactification, and in particular in a warped Calabi-Yau, take the same form
as in the standard Calabi-Yau case. Note, however, that in general
the forms $J$ and $\Omega$ are no longer calibrating forms in
$\cm_6$, since they are not closed.

\subsection{Wrapping D-Branes}

Let us now briefly discuss the various possibilities for type IIB D-branes wrapping $2n$-cycles of $\cm_6$. The main differences with the cases without fluxes arise due to the fact that the background selects a special ${\cal N}=1$ supersymmetry generator and branes have to be compatible with it. In addition, we now have an $H_3$ flux, which can give rise to novel effects.

\vspace*{.25cm}

{\bf Two-Cycles}

\vspace*{.125cm}

\noindent
This case applies to a  D1 instanton or a space-filling D5-brane.
The constraint \eqn{mmms2} forces the two cycle $\cs_2 \subset
\cm_6$ to be   holomorphic. The condition  (\ref{mmms1}) reads
\be
 J+ i \cf = i\, { \sqrt{|g + \cf|} \over  \sqrt{|g|}} \, d {\rm vol}_{\cs_{2}}.
 \label{n=1}
\ee
The only possible solution of this equation is to  consider  a
configuration where $\cf \raw \infty$ and $J|_{\cs_2} \raw 0$
simultaneously. Physically this corresponds to a situation where
the size of the two-cycle is shrunk to zero size while the  amount
of flux through the vanishing cycle  is scaled  in such a way that
the collapsed D5-brane has a non-vanishing tension, giving rise to
a fractional D(-1)/D3-brane. This type of branes are ubiquitous
near Calabi-Yau singularities such as orbifolds and conifolds, being
consistent supersymmetric branes in the corresponding flux backgrounds.

As evidenced from this analysis only fractional branes are
consistent with supersymmetry. This could have intuitively been
expected by noting that the branes have to be compatible with the
projection imposed on the Killing spinor of the background, which
is that of a regular D3-brane and recalling  that a finite size
space-filling D5-brane is not compatible with that projection.

 \vspace*{.25cm}

 {\bf Four-Cycles}

 \vspace*{.125cm}

 \noindent
Let us now turn to the case of a space-filling  D7-brane or a
D3-brane instanton wrapping a four-cycle on $\cm_6$. We expect this case to be more involved and interesting than the previous one since, as has been shown in the literature, background fluxes can lift D7-brane geometrical moduli \cite{gktt,gencali,ciu2,lmrs}. From the point of view of Euclidean D3-branes, these lifted moduli translate into a four-dimensional instanton with less zero modes than its (unwarped) Calabi-Yau analogue. As a result, there may be new D3-brane instantons contributing to the four-dimensional superpotential. We would expect such effects to be visible from the supersymmetry equations. 

The equations we derived demand that the four-cycle is a divisor $\cs_{4}$ which is holomorphically embedded in $\cm_6$. Equations \eqn{mmms1} and \eqn{mmms2} read in this case
 \bea
 \cf^{(2,0)} = \cf^{(0,2)} & = & 0 \,, \label{D71}\\
 J\wedge \cf & = & 0 \label{D72} \,,
 \eea
so we recover that $\cf$ must be a (real) primitive\footnote{By primitive we mean any $p$-form $\alpha$ such that $\alpha \wedge J = 0$.}
(1,1)-form on $\cs_4$.

This condition on $\cf$ can like-wise be written as the anti-selfduality condition on the divisor $*_4 \cf = - \cf$. The fact that supersymmetry requires
a specific polarization of $\cf$ can be physically motivated by the observation that an anti-selfdual field configuration on the worldvolume of a D7-brane induces D3-brane charge, which aligns with the supersymmetry broken by the background. Therefore, the supersymmetry conditions (\ref{D71}) and (\ref{D72}) mean nothing but that we need to consider D7-branes whose worldvolume bundles carry a D3-brane charge.

\vspace*{.25cm}

{\bf Wrapping the Calabi-Yau}

\vspace*{.125cm}

\noindent

The final case to consider is that of a space-filling D9-brane or a D5-brane wrapping the entire manifold $\cm_6$. Equations \eqn{mmms1} and \eqn{mmms2} reduce in this case to
 \bea
 \frac{1}{3!} \cf^3 - \frac{1}{2!} J^2 \wedge \cf & = & { \sqrt{|g + \cf|}  \over \sqrt{|g|}} \, d {\rm vol}_{\cm_{6}},
 \label{D91}\\
 \frac{1}{3!} J^3 - \frac{1}{2!} J \wedge \cf^2 & = & 0
 \label{D92}\\
 \cf^{(2,0)} & = & 0
 \label{D93}
 \eea
These equations have solutions when $\cm_6$ is non-compact or when $H_3=0$. If the model is compact, so that $H_3$ does not vanish in cohomology, $\cf$ cannot be globally well defined because it must satisfy the Bianchi identity $d\cf = H_3$. Thus, as is stands, the D-brane worldvolume theory is not consistent and suffers from a Freed-Witten anomaly \cite{fw,mms}. It would be interesting to
find out under what conditions a globally defined $\cf$ can be
constructed, perhaps by adding extra sources of $\cf$ along the proposal in \cite{cu}.

\section{The Moduli Space of D7-branes: Local Aspects}\label{local}

The goal of this section is to analyze the local structure of the
moduli space of supersymmetric D7-branes in a flux
compactification \eqn{warped0}. Specifically, we will assume that
a solution exists, and ask what the possible deformations of the
divisor $\cs_4 \subset \cm_6$ wrapped by the D7-brane are. We will
see that all geometric moduli are generically lifted
with respect to the fluxless case.

The NS-NS and RR background fluxes $H_3$ and $F_3$ may be combined
into the complexified three-form flux
\be
G_3 = F_3 - \tau H_3,
\label{G3}
\ee
where $\tau = C_0 + i/g_s$ is the type IIB axion-dilaton field,
which we take to be constant for simplicity. These fluxes must
satisfy the Bianchi identities $d H_3 = d F_3 = 0$, as well as the
proper quantization conditions on three-cycles of
$\cm_6$. $G_3$ must be ISD, namely $*_6 G_3 = i G_3$, and hence
harmonic. Supersymmetry further requires $G_3$ to be a primitive
form (i.e., such that $G_3 \wedge J =0$) of type (2,1)
\cite{gp}. Since $H_3=- \pim G_3 / \pim \tau$, it follows
that $H_3$ is a real three-form such that
\be
H_3 \in H^{(2,1)}_{p} \oplus H^{(1,2)}_{p} \,,
\ee
where the subindex `$p$' stands for primitive forms. A self-dual
RR five-form flux is also present, but its precise form will not
be needed here.

Consider now a supersymmetric D7-brane wrapping a four-cycle
$\cs_4$ of $\cm_6$, with a worldvolume $U(1)$ gauge field $\cf$.
By assumption, the four-cycle is holomorphic and $\cf$ is
anti-selfdual. We wish to determine the local moduli space for
this D7-brane, namely the continuous deformations that respect
these supersymmetry conditions. In the fluxless case, the moduli
space would be parametrized by\footnote{More precisely,
$\z^a$ and
$\xi^b$ should be thought of as massless chiral multiplets
transforming in the adjoint representation of the gauge group of
the D7-brane. They may or may not be true moduli, as there may
exist a non-trivial superpotential $W$ for these fields with no
quadratic terms.}
 \cite{jlD7}
\bea
{\rm Geometric\ moduli}: & & \z^a \quad \, a = 1, \dots, H^{(0,2)}
(\cs_4) \,,
\label{gmoduli}\\
{\rm Wilson\ line\ moduli}: & & \xi^b \quad \, b = 1, \dots, H^{(0,1)} (\cs_4)
\label{wlmoduli} \,.
\eea
The geometric moduli parametrize deformations that preserve the
holomorphicity of the cycle. We will now examine under what
circumstances such deformations can respect the anti-selfduality
condition on $\cf$.

We begin by recalling that in order for the gauge theory on the
D7-brane to be anomaly-free, the cohomology class in
$H^3(\cs_4)$ defined by the pull-back of $H_3$ must vanish
\cite{fw}; note that this condition is invariant under continuous
deformations of the cycle. Under these circumstances, the full
content of the Bianchi identity
\be
\label{eqF}
d \cf  =  H_3
\ee
is to imply that $\cf$ is a globally well defined form on $\cs_4$.
By the Hodge decomposition theorem, we may uniquely decompose it
as
\be
\cf = h + d \xi + d^\dagger \eta \,,
\label{decomp}
\ee
where $h$ is harmonic \cite{candelas}. The anti-selfduality
condition then translates into $d\xi = - *_4 d ^{\dagger} \eta$
and $h^\mt{SD} = 0$, where $h^\mt{SD}$ is the self-dual component
of $h$. One solution of the first equation is $\xi = *_4 \eta$,
which may always be achieved by a convenient choice of the
worldvolume gauge potential
$A$ in \eqn{f}.

We are thus left with a non-trivial constraint only on the
variation under a holomorphic deformation of the harmonic part of
$\cf$. To examine this, let us expand it as
\be
\cf^\mt{har} = \sum_{i = 1}^{h^{0,2}}\, a_i (\z)\, \a^i + c.c.
+ \sum_{j = 1}^{h^{1,1}_{p}}\, b_j (\z)\, \b^j
+  c (\z)\, \g \,,
\label{decompF}
\ee
where $\a^i$, $\b^j$ and $\g$ are a basis of harmonic two-forms
belonging to the middle cohomology of $\cs_4$ as shown in table
\ref{middle}. 
\TABLE{\renewcommand{\arraystretch}{1.2}
\begin{tabular}{ccc}
$[\bar{\a}^i]\ \in\  H^{2,0}$ & $[\b^j]\ \in\ H^{1,1}_{p}$ &
$[\a^i]\ \in\ H^{0,2}$ \\
& $[\g]\ \in\ H^{1,1}_{np}$  \\
\end{tabular}
\label{middle}
\caption{\small Middle cohomology of $\cs_4$ and harmonic representatives.}}
The fact that $\cs_4$ is a four-dimensional conformally K\"ahler submanifold implies that there is a unique cohomology class of harmonic (1,1) non-primitive forms, whose harmonic representative $\tilde{\g}$ is proportional to $J^\mt{CY}$. It also implies that $\a^i, \bar{\a}^i$ and $\g$ are self-dual, whereas $\beta^j$ are anti-selfdual. The coefficients $a^i$, $b^j$, and $c^j$ may depend on the positions $\z^a$ of the four-cycle, but not on the choice of Wilson lines $\xi^b$, since $\cf$ is independent of these. Without loss of generality, we take the initial $\cs_4$ to correspond to $\z=0$. Since this is supersymmetric by assumption, we have $a^i(0) = c(0)=0$. 

Let $\tilde{\a}_i$, $\tilde{\b}_j$ and $\tilde{\g}$ be sets of harmonic dual forms, such that
\be
\int_{\cs_4} \tilde{\a}_i \, \wedge \, \a^j = \d_i^j \sac
\int_{\cs_4} \tilde{\b}_i \, \wedge \, \b^j = \d_i^j \sac
\int_{\cs_4} \tilde{\g} \wedge \g = 1 \,.
\ee
Then
\be
a_i = \int_{\cs_4} \tilde{\a}_i \wedge \cf^\mt{har} =
\int_{\cs_4} \tilde{\a}_i \wedge \cf \,,
\label{a}
\ee
where in the last equation we used the fact that $\tilde{\a}_i$ is harmonic. In fact, the  $\tilde{\a}_i$ may be constructed as follows. An  infinitessimal deformation of $\cs_4$ that preserves its holomorphicity is specified by a holomorphic section $X$ of the normal bundle of $\cs_4$. Let $X_i$ be a basis of such sections. Then the pull-back of $\iota_{X_i} \Omega^\mt{CY}$ is a basis of harmonic (2,0) forms on $\cs_4$, and this map provides an isomorphism $\tilde{\a}_i = \iota_{X_i} \Omega^\mt{CY}$ between holomorphic deformations of $\cs_4$ and $H^{2,0}(\cs_4)$. In this basis, equation \eqn{a} becomes
\be
a_i (\cs_4)\, = \,  \int_{\cs_4} \iota_{X_i} \Om^\mt{CY}
\wedge \cf \,,
\label{ai}
\ee
which vanishes identically if the cycle is holomorphic and the supersymmetry condition \eqn{mmms2} is satisfied, as expected. Analogously, the condition (\ref{D72}) directly implies that the coefficient $c(\cs_4)$ vanishes.


Let us now determine how these coefficients vary as we deform the four-cycle. For any two four-cycles $\cs_4$ and $\cs_4'$ related by a continuous deformation there exists a five-chain $\Sigma_5$ such that $\partial \Sigma_5 =\cs_4' - \cs_4$. Stokes' theorem and the Bianchi identity \eqn{eqF} imply that
\be
a_i (\cs_4') - a_i(\cs_4) 
= \int_{\Sigma_5} \iota_{X_i} \Omega^\mt{CY} \wedge H_3 
+ d(\iota_{X_i} \Omega^\mt{CY})\wedge \cf\, ,
\label{deltaa}
\ee
where $\cf$ is now the natural extension of the worldvolume flux to the five chain $\Sigma_5$. Notice that (\ref{deltaa}) identically vanishes whenever $\cf$ comes from the pull-back of a closed (1,1) B-field.

The coefficients $b_j$ and $c$ are given by formulas analogous to \eqn{a} with $\tilde{\a}_i$ replaced by $\tilde{\b}_j$ and $\tilde{\g}$, respectively. Notice that Stokes' theorem now implies that $c$ is constant under holomorphic deformations, since
\be
c(\cs_4') - c(\cs_4) = \, \int_{\Sig_5} J^\mt{CY} \wedge H_3 =0
\label{changeJ}
\ee
by virtue of the facts that $\tilde{\g}$ can be taken to be the 
pull-back of $J^\mt{CY}$ and $H_3$ is primitive.

The coefficients $b_j$ will generically change under the deformation $\cs_4 \raw \cs_4'$, but recall that the forms $\b^j$ are anti-selfdual and hence compatible with supersymmetry. We thus conclude that the conditions for a holomorphic deformation of $\cs_4$ to preserve supersymmetry are
\bea \nonumber
a_1(\z^1,\dots,\z^{h^{0,2}}) &=& 0 \,, \\
\vdots
\label{submoduli}\\ \nonumber
a_{h^{0,2}}(\z^1,\dots,\z^{h^{0,2}}) &=& 0 \,.
\eea
We will see below that the $a_i$ are holomorphic functions of the deformations $\{\z^i\}$. 
Hence, (\ref{submoduli}) defines (locally) a complex submanifold within the moduli space 
of $\cs_4$ that will be denoted as $\cm^F(\cs_4) \subset \cm (\cs_4)$. Intuitively, 
the codimension of $\cm^F(\cs_4)$ inside $\cm (\cs_4)$ equals the number of geometrical 
moduli lifted by the presence of the flux $H_3$. Notice that (\ref{submoduli}) is a 
system of $h^{0,2} (\cs_4)$ equations for $h^{0,2} (\cs_4)$ unknowns. Therefore, 
generically the solution will consist of a set of isolated points, and so all the 
geometrical moduli of the D7-brane will be lifted.

The number of infinitessimal deformations around a given point is $h^{0,2}-r$, 
where $r$ is the rank of the matrix
\be
{\p a_i \over \p \z^j} =  
\int_{\cs_4} L_{X_j} \left(\iota_{X_i} \Om^\mt{CY} \wedge \cf \right)  
=  \int_{\cs_4} L_{X_j} \left( \iota_{X_i} \Om^\mt{CY}\right) \wedge \cf + \iota_{X_i}
\Om^\mt{CY} \wedge \iota_{X_j} H_3.
\label{derivative}
\ee
Here we have used the fact that $\z^i$ may be though as the components of a 
holomorphic section $X$ in a given basis $\{X_i\}$, namely $X=\z^i X_i$. 
Notice that 
$L_{X_j} \iota_{X_i} \Om^\mt{CY} = \iota_{[X_j,X_i]}\Omega+\iota_{X_i}d\iota_{X_j} \Omega$ 
is again a (2,0)-form, where $L_{X_j}$ stands for the Lie derivative with respect to 
the holomorphic deformation represented by $X_j$. Hence the r.h.s. of (\ref{derivative}) 
simplifies once we impose the supersymmetry condition (\ref{D71}), and we end up with
\be
{\p a_i \over \p \z^j} = 
\int_{\cs_4}  \iota_{X_i}
\Om^\mt{CY} \wedge \iota_{X_j} H_3  =  \frac{ig_s}{2}
\int_{\cs_4}  \iota_{X_i} \Om^\mt{CY} \wedge \iota_{X_j}
\bar{G}_3 \,.
\label{derivative2}
\ee
where the last equality follows from $H_3 = - \pim G_3/ \pim \tau$ and the fact that $G_3$ is a (2,1)-form. In particular, this implies that $(\p a_i/ \p \bar{\z}^{\bar{j}}) = 0$, so the coefficients $a_i$ are indeed holomorphic functions of $\{\z^i\}$.\footnote{Poincar\'e invariance of the ten-dimensional background along the four non-compact directions also allows a (0,3) component of $G_3$, which breaks supersymmetry. In this case the coefficients $a_i$ need no longer be holomorphic functions.}

From the viewpoint of the four-dimensional effective field theory,
the conditions for unbroken supersymmetry should arise from an
effective superpotential $\cw$ for the open string moduli. A
natural interpretation is to identify the equations
$a_i(\xi)=0$ with $\partial_i \cw =0$, since both are holomorphic
functions of the moduli. Note that these are indeed the equations
for unbroken supersymmetry if the potential is of no-scale type.
It would be interesting to verify this proposal in explicit
examples.

\section{The Moduli Space of D7-branes: Global Aspects}\label{global}

As we have shown, $\k$-symmetry provides us with a powerful tool to
analyze the moduli space of a D7-brane in a flux background. Our
analysis so far, however, has only provided us with a local
description of the moduli space. Indeed, we have seen that for a
D7-brane wrapping a supersymmetric four-cycle $\cs_4$, a background
flux $G_3$ reduces its geometrical moduli space to a subvariety
$\cm^F(\cs_4) \subset \cm(\cs_4)$. Generically all the geometrical
moduli are lifted, so $\cm^F(\cs_4)$ will locally be an isolated
point of $\cm(\cs_4)$. However,  the global structure of
$\cm^F(\cs_4)$ may be more involved, and it may consist of a
discrete set of points inside $\cm(\cs_4)$. One of the main
purposes of the present section is to show that this will be
usually the case.

It is clear that these global issues are of central interest for the statistical analysis of flux vacua in string theory \cite{counting}, once that open strings are introduced in such analysis. Indeed, we will argue that including D7-branes in flux compactifications opens up a new landscape of possibilities for the type IIB discretum. Moreover, this open string landscape is directly connected to the physics of $D=4$ non-Abelian gauge theories and chiral matter, so any statistical prediction derived from it should have a clear interpretation in terms of low energy particle physics.

\subsection{A toroidal example}

Before performing a general analysis, let us illustrate both local
and global features of the moduli space of D7-branes by
considering an explicit flux compactification. For simplicity, we
will consider a background such that constant fluxes and flat
space are solution of the supergravity equations of motion. The
simplest compact spaces of this kind are given by $\T^6$ or
toroidal orbifolds such as $\T^4/\ZZ_2 \times \T^2$, $\T^6/(\ZZ_2
\times \ZZ_2)$ \cite{kst,fp,tri03,blt,cu}. Since these backgrounds
are compact, the ansatz (\ref{warped0}) requires sources of
negative tension in the construction, such as O3$^-$-planes.
Hence, we need to further mod out these toroidal backgrounds by
$\Om^{ws}\car(-1)^{F_L}$, where $\Om^{ws}$ is the worldsheet
parity operator and $\car$ acts by $ (z^1,z^2,z^3) \mapsto
(-z^1,-z^2,-z^3)$. For the sake of definiteness, let us choose the
metric on $\T^6$ to be of the form $(\T^2)_1 \times (\T^2)_2
\times (\T^2)_3$, and with background fluxes
\bea
F_3 & = & 4\pi^2 \a'\, N \, \left(dx^1 \wedge dx^2 \wedge dy^3\,
+\, dy^1 \wedge dy^2 \wedge dy^3\right)
\label{fluxF}\\
H_3 & = & 4\pi^2 \a'\, N \, \left(dx^1 \wedge dx^2 \wedge dx^3\,
+\, dy^1 \wedge dy^2 \wedge dx^3\right)
\label{fluxH}
\eea
where $0 \leq x^i, y^i \leq 1$ and $N \in \ZZ$.\footnote{In
general, the integer $N$ cannot take an arbitrary value, but needs
to be of the form $N = n N_{\rm min}$. Here $n \in \ZZ$ and
$N_{\rm min} \in \NN$ depends on the homology of three-cycles in
${\T^6}/\G$, as well as on the O3-plane content on the quotient
space. For compactifications where only the usual O3$^{-}$-planes
appear, we have that $N_{\rm min} = 2$ for $\G = Id.$ \cite{fp}
and that $N_{\rm min} = 4,8$ for $\G = \ZZ_2 \times \ZZ_2$,
depending on the choice of discrete torsion \cite{blt,cu}. Roughly
speaking, $N_{\rm min}$ indicates the minimum amount of flux that
we can turn on.}

The fluxes (\ref{fluxF}), (\ref{fluxH}) generate a  superpotential
\be
W\, =\, \int \Om \wedge G_3\, \propto\, (1 + \tau_1 \tau_2) \cdot
(1 + \tau_3 \tau),
\label{superp}
\ee
$\tau_i$ being the complex structure of the $i^{th}$ $\T^2$ factor, and $\tau$ the complex dilaton-axion field. The supersymmetric minima of $W$ are given by
\be
\tau_1 \tau_2 = -1, \quad \quad \tau_3 \tau = -1
\label{minima}
\ee
which means that for this choice of fluxes only two combinations
of complex structure moduli/dilaton are stabilized.

We now want to introduce D7-branes in our compactification. As we
know, these D7-branes must wrap holomorphic four-cycles $\cs_4$ and
the pull-back $H_3|_{\cs_4}$ must be trivial in cohomology. The
simplest possibility is to consider a D7-brane wrapping two
complex dimensions $(\T^2)_j \times (\T^2)_k \subset (\T^2)_1
\times (\T^2)_2 \times (\T^2)_3$ and being pointlike in
$(\T^2)_i$, $j \neq i \neq k$. Following the type IIB model
building literature, we will name such D7-brane as D7$_i$ (see
fig.\ref{Zeeman} below for an illustration of a D7$_1$-brane).
Each of these D7-branes has the topology of $\T^4$, and hence its
middle cohomology is given by $h^{2,0} = h^{0,2} = 1$,
$h^{1,1}_{p} = 3$  and $h^{1,1}_{np} = 1$. As a consequence, each
will have one   geometrical modulus, which we will denote by
$\z^i$, which  parametrizes the motion along  $(\T^2)_i$.

The presence of a non-trivial $H_3$ on $\T^6$ gives rise to a
non-closed B-field, which eventually will contribute to the gauge
invariant two-form $\cf$ on the D7-branes. Let us see how this
happens for the D7$_i$-branes in our example. We will first
consider a D7$_1$-brane. A suitable choice of gauge for the
B-field in this case is given by
\be
B = 4\pi^2 \a'\, N\, \left(x^1 dx^2 \wedge dx^3 + y^1 dy^2
\wedge dx^3 \right)
\label{BD71}
\ee
which is well-defined on the patch $(\left \{0 \leq x^1 < 1\}
\times \{0 \leq y^1 < 1\} \right) \times (\T^2)_2 \times (\T^2)_3$
inside $(\T^2)_1 \times (\T^2)_2 \times (\T^2)_3$. Notice that
this choice satisfies $B_{\mu\nu} (-x^i,-y^i) = - B_{\mu\nu}
(x^i,y^i)$, compatible with the action $\car B = -B$ of the
orientifold group on the B-field. In addition, (\ref{BD71}) is
globally well-defined over the four-cycle wrapped by the
D7$_1$-brane, i.e., $\{\z^1\} \times (\T^2)_2 \times (\T^2)_3$.
Hence, in order to compute its pull-back we only need to consider
this single patch, and we obtain
\bea \label{pullD71}
B|_{D7_1} & = & 4\pi^2 \a'\, N\, \left(x^1 dx^2 \wedge dx^3 +
y^1 dy^2 \wedge dx^3 \right) \\ \nonumber & = & 4\pi^2 \a'\, N\,
\frac{i}{2} \left( \frac{\tau_2}{\pim \tau_2} z^1
d\bar{z}^{\bar{2}} + \frac{\tau_1}{\pim \tau_1} \bar{z}^{\bar{1}}
dz^2 \right) \wedge {\pim (\bar{\tau}_3 dz^3) \over \pim \tau^3}
\eea
where we have made use of the relations (\ref{minima}). Notice
that $B|_{D7_1}$ is a  harmonic form, which is consistent with
the fact that $H_3|_{D7_1}$ identically vanishes.

It is it easy to see that the $(0,2)$ component of $B|_{D7_1}$ depends holomorphically on $z^1$, which is nothing but the D7$_1$-modulus $\z^1$. Indeed, in terms of the general expression (\ref{decompF}), (\ref{pullD71}) translates into
\bea \label{a1ex}
a_1 & = & -  (4\pi^2\a')\, 4 N\, \z^1\, {\tau_2} {\tau_3} \\
b_1 & = &  (4\pi^2\a')\, 4 N\, \preal \left(\z^1\, {\tau_2} {\bar{\tau}_3} \right)\\
b_2 & = & -  (4\pi^2\a')\, 4 N\, \pim \left(\z^1\, {\tau_2} {\bar{\tau}_3} \right)\\
b_3 & = & 0 \\
c & = & 0
\eea
where we have identified $\z^1$ with $z^1$. Here $b_1$ and $b_2$
correspond to the coefficients of the real forms $\preal (dz^2
\wedge d\bar{z}^{\bar{3}})/ \pim \tau_2 \pim \tau_3$ and $\pim
(dz^2 \wedge d\bar{z}^{\bar{3}})/ \pim \tau_2 \pim \tau_3$,
respectively. Notice that the fact that $c=0$ was expected from
the general results of Section \ref{local}.

It is now clear that $B$ will induce a non-vanishing $(2,0) +
(0,2)$ component on the D7$_1$-brane, unless we choose the
specific location $\z^1 = 0$ for it. We then
conclude that the geometrical modulus $\z^1$ of D7$_1$-branes is
lifted after we introduce the fluxes (\ref{fluxF}) and
(\ref{fluxH}).

A similar result is obtained for D7$_2$-branes in the same
background. The case of D7$_3$-branes, however, is different.
Indeed, the pull-back of the B-field on $(\T^2)_1 \times (\T^2)_2
\times \{\z^3\}$ is given by
\bea \nonumber
B|_{D7_3} & = & 4\pi^2 \a'\, N\, \frac{\pim \bar{\tau}_3\z^3}{\pim \bar{\tau}_3} \left(dx^1 \wedge dx^2 + dy^1 \wedge dy^2 \right) \\
& = & 4\pi^2 \a'\, N\, \frac{\pim \bar{\tau}_3\z^3}{2i \pim
\bar{\tau}_3} \left( \frac{\tau_2}{\pim \tau_2} dz^1 \wedge
d\bar{z}^{\bar{2}} + \frac{\tau_1}{\pim \tau_1} d\bar{z}^{\bar{1}}
\wedge dz^2 \right),
\label{pullD73}
\eea
so that we find that $B|_{D7_3}$ defines a primitive (1,1)-form
regardless of the value of $\z^3$. That is, $H_3$ induces a
supersymmetric B-field on $D7_3$, independently of its position
and, as a result, no geometrical moduli are stabilized for
D7$_3$-branes.

Notice that these results match exactly with those obtained in
\cite{ciu2}, where  the effect of  $G_3$ flux  was computed on the
effective theory of a D7-brane wrapping a $\T^4$. Again using
(\ref{minima}), we can write the type IIB flux $G_3 = F_3 - \tau
H_3$ obtained from (\ref{fluxF}) and (\ref{fluxH}) as
\be
G_3 \, = \, 4\pi^2 \a'\, N\, \frac{1}{2i} \left(
\frac{\tau_2}{\pim \tau_2} dz^1 \wedge d\bar{z}^{\bar{2}} +
\frac{\tau_1}{\pim \tau_1} d\bar{z}^{\bar{1}} \wedge dz^2 \right)
\wedge {dz^3 \over \tau_3}
\label{fluxG}
\ee
and it is easy to see that the nonvanishing components of $G_3$ in terms of $SU(3)$ irreducible representations are given by $S_{\bar{1}\bar{1}}$ and $S_{\bar{2}\bar{2}}$, as defined in \cite{gp}. By the results of \cite{ciu2}, these are the flux components which give mass to the geometrical moduli of D7$_1$ and D7$_2$-branes, respectively, whereas they do not affect the D7$_3$-brane modulus.

In fact, one can be more precise and compute the supersymmetric masses of the D7-brane moduli $\z^1$ and $\z^2$, again using the techniques of Section \ref{local}. In order not to digress from the present discussion, we will postpone such analysis to Section \ref{app}.

\vspace*{.25cm}

{\bf A discrete set of $\cn=1$ D7-branes}

\vspace*{.125cm}

So far we have used a toroidal example to illustrate the general discussion of Section \ref{local}. However, being a compact model it turns out to be also useful to describe global aspects of the D7-brane moduli space. For instance, since we found that D7$_1$-brane geometrical moduli are lifted, one may wonder if there is only one or several values of $\z^1$ allowed by supersymmetry. We know that there is at least one, given by  $\z^1 = 0$, but in general there may be a discrete set of solutions.

Let us illustrate this fact with the choice of fluxes above.
Notice that in the discussion above we have implicitly assumed
that $\cf_{D7} = B|_{D7}$. That is, we have set $F = dA = 0$ on
(\ref{f}). In general, we must consider the gauge invariant
quantity $\cf_{D7_1} = 2\pi \a' F_{D7_1} + B|_{D7_1}$, whose
components in this case are given by
\bea
B|_{D7_1} & = & 4\pi^2 \a'\, N\, \left(\z^1_x\, dx^2 \wedge dx^3
+ \z^1_y\, dy^2 \wedge dx^3 \right)
\label{D71B}\\
F_{D7_1} & = & 2\pi\, \left(n_1\, dx^2 \wedge dx^3 + n_2\, dy^2
\wedge dx^3 + \dots\right)
\eea
where have parametrized the moduli space of the D7$_1$-brane as
$\z^1 = \z^1_x + \tau_1 \z^1_y$ for convenience. Here $n_1, n_2$
are integer numbers and `\dots' stands for extra components of $F$
which will not be directly relevant in the following, and we will
assume them to vanish. Notice that both $F_{D7_1}$ and $B|_{D7_1}$
are harmonic forms on $\{\z^1\} \times (\T^2)_2 \times (\T^2)_3$,
but that only $F_{D7_1}$ needs to satisfy the Dirac quantization
condition.

Now, in order to look for supersymmetric solutions for a
D7$_1$-brane, we can search for choices of $B|_{D7_1}$ and
$F_{D7_1}$ which give us a vanishing
$\cf_{D7_1}$.\footnote{Actually, we should be looking for
combinations such that $\cf_{D7_1}^{(2,0)}$ vanishes, which is a
milder condition. However, it does not give new solutions in the
case at hand.} Since $\cf_{D7_1}$ is given by the combination
(\ref{f}), it is clear that such cancellation will not be possible
for arbitrary values of $(\z_x^1, \z_y^1)$. The only possibilities
are given by
\be
\begin{array}{c}
\z_x^1 N\, =\, -n_1 \, \in\, \ZZ \\
\z_y^1 N\, =\, -n_2 \, \in\, \ZZ
\end{array}
\label{solns}
\ee
which indeed give us a discretum of possibilities. Each of these
represents a D7$_1$-brane such that $\cf_{D7_1}$ vanishes, and is
hence trivially supersymmetric.

As an example, let us consider the background $\T^6/\Om^{ws} \car
(-1)^{F_L}$, such that there are 64 O3$^-$-planes in our
compactification. The positions of such O3-planes, or more
precisely their projection onto $(\T^2)_1$, are given by $(\z_x^1,
\z_y^1) = \{(0,0), (0,\oh), (\oh,0), (\oh,\oh)\}$. Each of these
points is actually equivalent from the point of view of the
compactification. In particular, since we have obtained that the
origin $(\z_x^1, \z_y^1) = (0,0)$ is a supersymmetric location for
a D7$_1$-brane, so must be the equivalent points $(0,\oh),
(\oh,0), (\oh,\oh)$. Now, by looking at the general solution
(\ref{solns}) we see that these four points will be supersymmetric
locations if and only if $N \in 2\ZZ$. This may seem quite a
restrictive set of fluxes at first sight but, as pointed out in
\cite{fp}, $N \in 2\ZZ$ are indeed the allowed choices of flux for
this orientifold background. It is amusing that we have rephrased
the flux quantization conditions of \cite{fp} in terms of
translational symmetries of the background $\T^6/\Om^{ws} \car
(-1)^{F_L}$.

\subsection{The open string landscape}

The toroidal example above teaches us an interesting lesson. As we
have seen in Section \ref{local}, the geometrical moduli of a
D7-brane are lifted by the presence of the background flux. This
reduces its geometrical moduli space to a subvariety $\cm^F(D7)
\subset \cm(D7)$ which, generically, is
zero-dimensional.\footnote{Notice that in all this discussion we
are neglecting the moduli space of Wilson lines on each D7-brane.
Since in principle they are not affected by the background flux
$G_3$, their contribution to the moduli space of the D7-brane
would remain untouched.} Locally this zero dimensional space looks
like a point, but globally this need not be the case. Indeed,
above we have shown that the former moduli space of a
D7$_1$-brane, which is nothing but $\cm(D7_1) = (\T^2)_1$, is now
replaced by a grid of points or two-dimensional lattice
$\cm^F(D7_1) = \Lam_2$, as illustrated in figure \ref{Zeeman}.

\EPSFIGURE{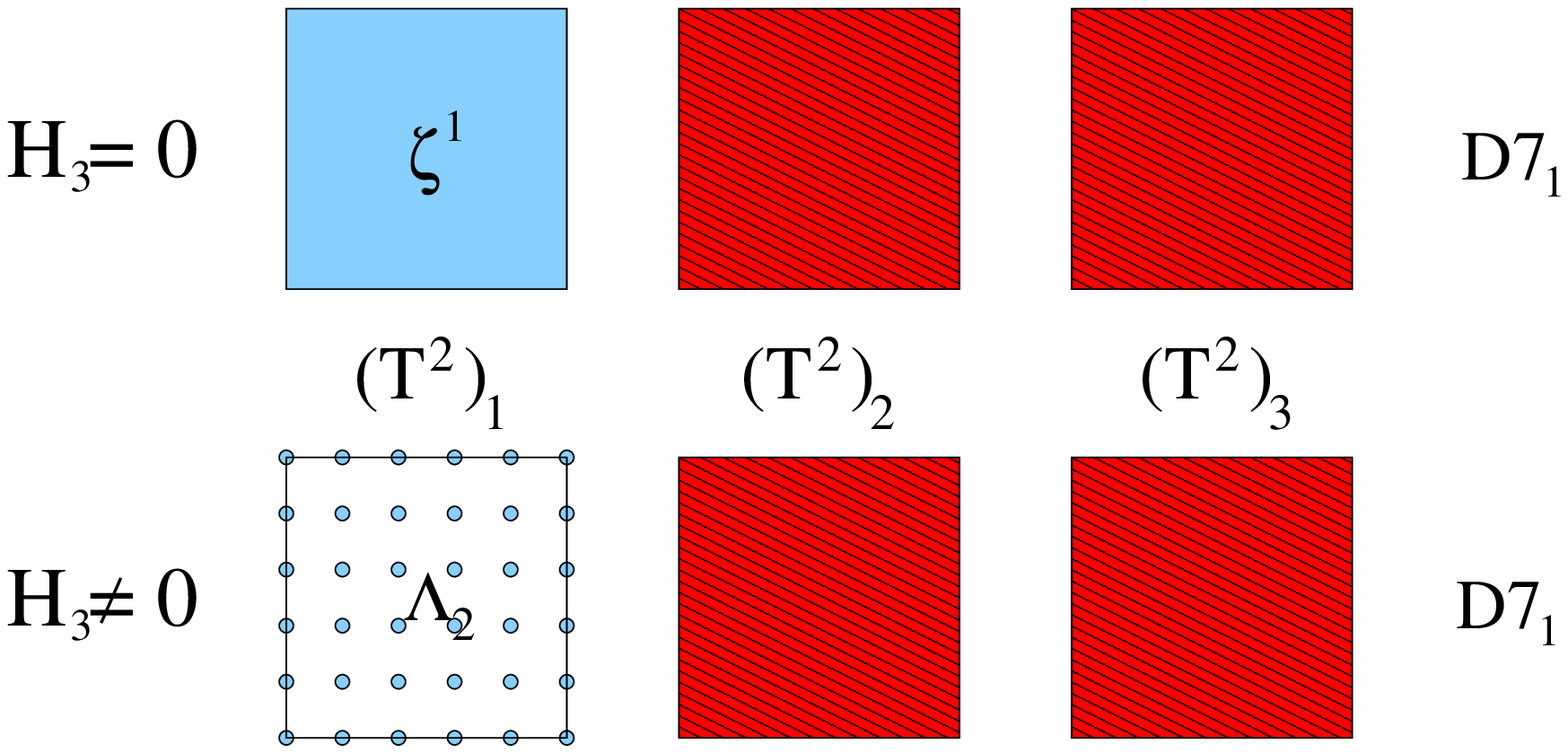, width=4.5in}
{\label{Zeeman} Open string landscape of a D7$_1$-brane. Before
the flux $H_3$ is introduced, the geometrical moduli space of a
D7$_1$-brane is the $\T^2$ where it is point-like, i.e., $(\T^2)_1$.
After the flux $H_3$ is turned on, this geometrical moduli space
is restricted to a lattice of points $\Lam_2 \subset (\T^2)_1$.}

This lattice of supersymmetric D7$_1$-branes is given by
(\ref{solns}), which can be expressed as
\be
N \z^1\, = \, m_x + \tau_1 m_y, \quad m_x, m_y \in \ZZ
\label{cpxsolns}
\ee
and which generalizes the previous solution $\z^1 = 0$, obtained
from a local analysis. Since $\z^1$ parametrizes the moduli space
of D7$_1$-branes, the lattice $\Lam_2$ can be expressed as
\be
\Lam_2 \, = \, \left\{ \left. {m_x + \tau_1 m_y \over N} \right| m_x, m_y \in \ZZ\right\}/ \Lam \quad \subset \quad (\T^2)_1
\label{lattice}
\ee
where $\Lam = \{n_x + \tau_1 n_y | n_x, n_y \in \ZZ \}$ is the
usual lattice of identifications that compactifies $\CC$ to
$(\T^2)_1$.

We can now attempt to extend these results to more general flux
compactifications. Let us consider type IIB compactified on
$\cm_6$, and containing a supersymmetric D7-brane wrapped on
$\cs_4$. In absence of fluxes or torsion, the geometrical moduli
space of such D7-brane would be given by $\cm(\cs_4)$,
parametrized by $\{\z^1, \dots, \z^{h^{0,2}}\}$.\footnote{In the
following, we are implicitly assuming that $\cm(\cs_4)$ is a
smooth manifold of definite dimension in order to simplify our
discussion. Our statements should be easy to generalize to more
involved cases.} When we introduce the background flux $G_3$, the
geometrical moduli space of the D7-brane should be reduced to
$\cm^F(\cs_4)$. Locally, this reduced moduli space would look like
a submanifold $\cm^f(\cs_4)$ defined by imposing the equations
(\ref{submoduli}) on the former moduli space. Globally, however,
we would expect a lattice of such $\cm^f(\cs_4)$'s. Schematically,
if the fluxes stabilize $q$ complex geometrical moduli of the
D7-brane we would expect something of the form
\be
\cm^F(\cs_4)\, = \, \Lam_{2q} \times \cm^f(\cs_4)
\label{open}
\ee
where $\cm^f(\cs_4)$ has complex dimension $h^{0,2}-q$, and is
transverse to the $2q$-dimensional lattice $\Lam_{2q} \subset
\cm(\cs_4)$.\footnote{More likely, (\ref{open}) would not be a
direct product, but the geometry and topology of $\cm^f(\cs_4)$
would vary for each lattice point of $\Lam_{2q}$.} For a generic
choices of fluxes, we would expect all the geometrical moduli of
the D7-brane to be lifted, so that $q=h^{0,2}$ and
$\cm^F(\cs_4) = \Lam_{2h^{0,2}}$ is a lattice of points inside
$\cm(\cs_4)$.

Whereas the above picture works at the intuitively level, it is important to realize that the D-brane supersymmetry conditions provide us with a precise definition of the new moduli space $\cm^F(\cs_4)$. In the present context this definition would read
\be
\cm^F(\cs_4)\, = \, \left\{ \cs_4 \in [\cs_4]\, \Big{|}\, B|_{\cs_4} \in A^{(1,1)}_p (\cs_4,{\RR}) \oplus  H^2_p (\cs_4,{\ZZ}) \right\},
\label{open2}
\ee
which means that $B|_{\cs_4}$ can either be a primitive real (1,1)-form of $\cs_4$ ($A^{(1,1)}_p (\cs_4,{\RR})$), a integral primitive two-form ($H^2_p (\cs_4,{\ZZ})$), or a combination of both. Intuitively, if $B|_{\cs_4}$ is (1,1) and primitive it does not break supersymmetry, and if it has a component on $H^2_p (\cs_4,{\ZZ})$ it can be cancelled with the appropriate gauge bundle $F= dA$ such that $\cf = 2\pi \a' F + B$ is again a real primitive (1,1)-form.

This definition turns out to be quite useful in the generic case
where $\cm^F(\cs_4)$ is the $2h^{(0,2)}$-dimensional lattice
$\Lam_{2h^{0,2}(\cs_4)}$. Indeed, to each point $l \in
\Lam_{2h^{0,2}(\cs_4)}$ we can now associate an element of $H^2_p
(\cs_4,{\ZZ})/(H_p^{(1,1)} (\cs_4,{\RR}) \cap H^2 (\cs_4,\ZZ))$,
by forgetting about the always supersymmetric $(1,1)_p$ component
of $B|_{\cs_4}$.

Let us now see how this new moduli space $\cm^F(\cs_4)$ and its
quantum numbers affect the spectrum of supersymmetric D7-branes.
The BPS D7-branes with $U(1)$ bundles can be classified in terms of
the following topological quantities
\begin{itemize}

\item[i)] $[\cs_4]\, \in\, H_4(\cm_6,\ZZ)$, such that $\cs_4$ is holomorphic and $[H_3|_{\cs_4}]$ is trivial.

\item[ii)] $[F]\, \in\, H_p^{(1,1)} (\cs_4,{\RR}) \cap H^2 (\cs_4,\ZZ)$

\item[iii)] $\Lam_{2h^{0,2}(\cs_4)}$

\end{itemize}
and to each topological sector we should attach the moduli space of $H^{0,1}(\cs_4)$ Wilson lines.

Notice that the topological charges $i)$ and $ii)$ are already present in standard Calabi-Yau compactifications with closed B-field. Indeed, $i)$ corresponds to the homology class $[\cs_4] \in H_4(\cm_6,\ZZ)$ of the four-cycle $\cs_4 \subset \cm_6$ that the D7-brane is wrapping, and $ii)$ to the Chern numbers $[F] \in H^2(\cs_4,\ZZ)$ of the $U(1)$ gauge bundle field strength $F=dA$. On the other hand, $\Lam_{2h^{0,2}(\cs_4)}$ is a new topological sector which arises after we introduce a background flux $G_3$, and that can be identified with a subset of $H^2_p (\cs_4,{\ZZ})/(H_p^{(1,1)} (\cs_4,{\RR}) \cap H^2 (\cs_4,\ZZ))$.

It is important to notice that, in general, two D7-branes D7$_\a$ and D7$_\b$ in two different sectors of $i)$ and $ii)$ can be continuously connected to a third D7-brane D7$_\g$ different from the previous ones. This can be done by simply giving a vev to the open strings living in the sector $\a\b$, triggering the process of D-brane recombination $D7_\a + D7_\b \raw D7_\g$. Hence, in a given construction many of the vacua which differ by the open string quantum numbers $i)$ and $ii)$ are in fact connected by a continuous deformation. In addition, in some cases the final D7-brane is not described by a $U(1)$ bundle but by a more general $U(N)$ bundle, and such BPS D-brane is not even captured by the classification above.

One may wonder if something analogous could happen for the
discretum of D7-brane positions $\Lam_{2h^{0,2}(\cs_4)}$. In
principle, since we have given a mass to the open strings that
connect every two points of $\Lam_{2h^{0,2}(\cs_4)}$, these fields
cannot acquire a vev and each point of
$\Lam_{2h^{0,2}(\cs_4)}$ labels a truly disconnected vacuum. On
the other hand, since our supersymmetry equations are based on the
hypothesis of D7-branes with $U(1)$ bundles, new supersymmetric
solutions may arise when we consider $\cf$ being the curvature of
a $U(N)$ bundle. Studying the spectrum of such D7-branes with $U(N)$
bundles is beyond the scope of this paper. However, the
computations performed in \cite{ciu2} for some simple cases
suggest that the present results will generalize more or less
directly to this more involved case.

Finally, one may wonder about the number of points contained in
the lattice $\Lam_{2h^{0,2}(\cs_4)}$, which is a question directly
relevant for counting the number of open string vacua in a given
compactification.\footnote{Notice that the D-brane statistical
analysis of ref.\cite{bghlw} deals exclusively with the
topological sectors $i)$ and $ii)$, which can be connected to each
other by giving vevs to fields. To the best of our knowledge, 
the effect of the disconnected, discrete positions of D7-branes 
has not been taken into account in the literature.}  
Presumably, the number of points in
$\Lam_{2h^{0,2}(\cs_4)}$ will grow exponentially with
$h^{0,2}(\cs_4)$, since this is the dimension of the
lattice. For fixed $\cs_4$ and $h^{0,2}(\cs_4)$, we also expect the
number of points of $\Lam_{2h^{0,2}(\cs_4)}$ to grow bigger for
larger flux quanta, since then $B|_{\cs_4}$ would vary more
quickly on $\cm(\cs_4)$ and (\ref{open2}) would have more points.
We indeed find such behavior for the toroidal example above. More
precisely, from (\ref{solns}) we see that the amount of allowed
locations for a D7-brane grows as $N^2$, which is proportional to
the D3-brane tadpole introduced by the background flux
$L=\int_{\cm_6} \frac{g_s}{2} G_3 \wedge \bar{G}_3$. It would be
interesting to see how this fact generalizes to more involved
geometries. This suggests that the the number of open
string vacua scales like 
\be
{\cal N}_{open}\sim L^{h^{0,2}}, 
\ee
where $L$ is the tadpole charge.

\section{Applications}\label{app}

A geometrical understanding of D-brane moduli lifting and of the
open string discretum, as the one provided via $\k$-symmetry
above, should in principle give us new perspectives into some
other issues related to D-branes and flux compactifications. The
purpose of this section is to illustrate this fact by means of
several direct applications of the results obtained above. We will
first comment on the relevance
of our results for the zero mode counting of type IIB D3-brane
instantons. We will also explain how this geometrical framework of
moduli stabilization may provide an alternative way of computing
D-brane soft terms, as the ones that develop on the D7-brane
effective theory in the presence of fluxes. Finally, we will point
out some possible applications of our results to D7-brane model
building.

\subsection{Zero mode counting of D3-brane instantons}\label{zero}

In the type IIB flux compactifications considered in this paper, the form of the perturbative superpotential leaves untouched the K\"ahler moduli of the Calabi-Yau. In order to get a rigid model, where no parameter can be continuously tuned, one must find a mechanism for fixing them. The proposal of \cite{kklt} is that these moduli are fixed  by  a non-perturbative superpotential\footnote{If a suitable matter content is present on the D7-branes, one can also generate a non-perturbative superpotential by gaugino condensation.} generated by D-brane instantons.

The study of non-perturbative contributions in M/F-theory was initiated by Witten \cite{witten} for backgrounds without flux. \cite{barren,tri05,saulina,kkt,bm} have revisited this important problem in the context where fluxes are present. In our work, we have found that open string moduli are generically lifted in the presence of fluxes. By supersymmetry, we expect  that the corresponding fermionic zero modes also get lifted. In the case without fluxes the number of geometric zero modes is given by $h^{0,2}$, where   $h^{0,2}$ is the Betti number of the four cycle that the instanton wraps. In the case with fluxes, the number of zero modes  reduces  to $h^{0,2}-r$, where $r$ is the rank of the matrix defined in (\ref{derivative}). Given that there are four-cycles that do not contribute to the superpotential in the case without flux because they have too many fermion zero modes, we expect that  new contributions to the non-perturbative superpotential will arise  in the case with fluxes. It would be interesting to analyze the problem in more detail.

\subsection{Computing soft terms geometrically}

Besides providing a general understanding of D7-brane moduli lifting, our techniques could also apply to compute the actual masses induced by the fluxes on the D7-brane worldvolume fields. A microscopic analysis of such effect was performed in \cite{ciu2}, where the soft SUSY-breaking terms induced on a D7-brane were computed via dimensional reduction of the Myers' action. In order to illustrate the use of our approach in computing soft terms, let us reproduce some of the results in \cite{ciu2}.

In principle, one should be able to compute the soft terms induced in the D7-brane bosonic sector by simply looking at the variation of its Lagrangian density. Certain deformations $\z^i$ of the D7-brane would increase its energy, and this should be understood as a mass for the bosonic field $\Phi^i$ that corresponds to this deformation. In order to make contact with the computations in \cite{ciu2}, let us consider a D7-brane wrapping $\cs_4$, such that the pull-back $H_3|_{\cs_4}$ vanishes identically and $\cf=0$. By deforming $\cs_4 \raw \cs_4'$, $\cf$ may become a non-vanishing self-dual two-form. Then, after the computations in the Appendix, one can see that the Lagrangian density increases by
\be
\d \cl = \int_{\cs_4'} g_s^{-1} \D^{-2} \cf^2 = 2 \sum_k |a_k|^2 \int_{\cs_4'} g_s^{-1} \D^{-2} \a^k \bar{\a}^k
\label{inclag}
\ee
where we have decomposed $\cf$ in terms of its self-dual components $\sum a_k \a^k + {\rm c.c.}$. The mass matrix for the geometrical deformations $\{ \z^i \}$ are then given by the second derivative of $\cl$, that is
\be
M^2_{i\bar{j}}\, =\, {\p \cl \over \p \z^i \p
\bar{\z}^{\bar{j}}}\, =\, 2 \sum_k {\p a_k \over \p \z^i}
\overline{\p a_k \over \p \z^j}  \int_{\cs_4} g_s^{-1} \D^{-2}
\a^k \bar{\a}^k
\label{masslag}
\ee
and where the derivatives $\p a_k/ \p \z^i$ are given by
(\ref{derivative}).

In order to simplify this expression, we can assume an homogeneous
warp factor $\D$ and a constant dilaton, just as considered in
\cite{ciu2}. We can then factorize the integral in
(\ref{masslag}), and the mass matrix $m^2_{i\bar{j}}$ becomes a
simple quadratic function on the derivatives $\p a_k/ \p \z^i$.

A particular simple case corresponds to a D7$_1$-brane on
$(\T^2)^3$, just as in the toroidal example of the previous
section. In this case the only geometric deformation is given by
the position on the first $\T^2$ factor $\z^1$, and the
corresponding (0,2) form is given by $\iota_{\z^1} \Om^{\bf CY} =
dz^2 \wedge dz^3$. In the presence of a (2,1) primitive flux
$G_3$, the derivative of $a_i$ is given by
\be
{\p a_1 \over \p \z^1}\, = \, \frac{ig_s}{2} \int_{(\T^2)_2 \times
(\T^2)_3}  \iota_{\z^1} \Om^{\bf CY} \wedge \iota_{\z^1}
\bar{G}_3\, = \,  \frac{ig_s}{2} \overline{G_{\bar{1}23}}
\label{der1}
\ee
after the fields and forms have been conveniently normalized.
Notice that the component $G_{\bar{1}23}$ is nothing but $\oh
S_{\bar{1}\bar{1}}$ in terms of $SU(3)$ irreducible representations.
We then obtain that the mass of the adjoint scalar $\Phi^1$
corresponding to the D7$_1$-brane deformation $\z^1$ is given by
\be
m^2_{1\bar{1}} \, = \, \frac{g_s}{8} |S_{1\bar{1}}|^2
\label{mass1}
\ee
reproducing the result of \cite{ciu2}.\footnote{There is
actually a discrepancy in a factor of 9, due to a
B-field gauge choice taken in \cite{ciu2} which is not 
globally well defined on the D7-brane worldvolume.}

Although we have derived the mass of the geometrical modulus
$\Phi^1$ in a very simple case of the toroidal D7$_1$-brane, in
principle the formula (\ref{masslag}) can be applied to more
involved geometries, with more than one geometric modulus and even a
non-homogeneous warp factor. In addition, one should be able to
derive the same kind of formulae for the more general case where
$\cf$ does not vanish. It would be very interesting to perform
such computation and compare it to the results in \cite{lrs,lmrs},
where general soft-terms were computed by using effective action
techniques.

\subsection{D7-brane model building}

Let us now discuss the relevance of our results to the construction of actual string models involving fluxes and D-branes. The fact that all D7-brane geometric moduli are typically lifted, and that there is a discrete set of choices for the vev's of these former moduli, will clearly open new possibilities for type IIB model building. Before addressing which these new possibilities may be, however, it proves useful to describe specific examples of BPS D7-branes in this class of compactifications.

In order to find such examples, let us first briefly recall some
general features of the type IIB vacua under study. As already
mentioned, a flux compactification yielding D=4 Poincar\'e
invariance needs the presence of negative tension objects like
O3$^-$-planes. The way to obtain O3-planes is to mod out a type
IIB supergravity background by $\Om^{ws} \car (-1)^{F_L}$. Here
$\Om^{ws}$ is the usual orientation reversal of the string
world-sheet, whereas $F_L$ is the space-time fermion number in the
left-moving sector. Finally, $\car$ is an homolorphic involution
which is also an isometry of the compact manifold $\cm_6$, and
which acts on the bispinor forms as \cite{ori}
\be
\car J\, = \, J, \quad \quad \car \Om\, = \, - \Om.
\label{actionR}
\ee
and on the background fields as
\be
\car B\, = \, - B, \quad \quad \car C_2\, = \, - C_2
\label{actionRB}
\ee
Finally, the background field strengths $H_3 = d B$ and $F_3 = d
C_2$ must be a sum of harmonic three-forms which are odd under the
action of $\car$, just like $\Om$.

Let us now consider a D7-brane wrapping a four-cycle $\cs_4$ invariant under $\car$, that is, $\car \cs_4 = \cs_4$. This may happen, e.g., when a D7-brane intersects an O3-plane. Let us assume that the pull-back $H_3|_{\cs_4}$ is trivial in cohomology, and that every (2,0)-form $\a_2 \in H^{(2,0)}(\cs_4)$ is even
under the action of $\car|_{\cs_4}$. It is then easy to see that such D7-brane satisfies all the supersymmetry conditions. Because $\cs_4$ is fixed by $\car$ it is automatically a complex submanifold of $\cm_6$. In addition, since $B$ is odd under the action of $\car$ and any $(2,0)$-form $\a_2$ and $J^\mt{CY}$ are even, the integrals $\int_{\cs_4} \a_2 \wedge B$ and $\int_{\cs_4} J^\mt{CY} \wedge B$ vanish. But these are exactly the conditions for the coefficients $a_i$ and $c$ in (\ref{decompF}) to vanish, which is equivalent to require that a D7-brane is supersymmetric.

A particular case of the above consists in placing a D7-brane on
top of an O7-plane. That is, we wrap the D7-brane around the
four-cycle $\cs_{O7}$ which is a fixed point locus of $\car$ (i.e.,
$\car z_0 = z_0\ \forall z_0 \in \cs_{O7}$). In this case we can
drop the mild conditions assumed above. Because $H_3$ is an odd
three-form under the action of $\car$, it is easy to see that the
pull-back $H_3|_{\cs_{O7}}$ vanishes, not only in cohomology but
identically. This is consistent with the fact that, due to
(\ref{actionRB}), the field $B$ must vanish on $\cs_{O7}$. Since
$B$ vanishes, we can always choose $\cf =0$, which clearly
satisfies all the supersymmetry requirements.

To summarize, we find that all the additional supersymmetry
conditions introduced by type IIB three-form fluxes are automatically
satisfied when D7-branes are placed on top of an
O7-plane,\footnote{That $\cs_{O7}$ defines a supersymmetric
four-cycle of $\cm_6$ was somehow to be expected, since an $O7$-plane
is part of the closed string/supergravity background, and our
working hypothesis in this paper is that such background preserves
$D=4$ $\cn=1$ supersymmetry.} and (under some mild assumptions)
when they intersect an O3-plane. The $\cn=1$ D7-branes are then
characterized by the same choices of $\cf$ as if the background
flux $G_3$ was not present.

Let us now consider which class of D7-branes may be more suitable
for constructing viable models of particle physics and/or
cosmology. In principle, a very interesting possibility consist of
D7-branes wrapping a four-cycle $\cs_4$ such that $H^{0,1}(\cs_4) = 0$.
Indeed, by Poincar\'e duality $H^3(\cs_4) = 0$, and any such
D7-brane is automatically free of Freed-Witten anomalies. In
addition, the moduli space of Wilson lines is zero dimensional, so
the only possible moduli are the geometric deformations of
$\cs_4$. However, we have seen that these moduli are generically
lifted by the presence of the background flux $G_3$ so, at the end
of the day, this gauge sector of the theory is free of open string
moduli.

The absence of D7-brane moduli translates in a $U(N)$ low energy theory without massless fields in the adjoint.\footnote{Strictly speaking, the absence of massless adjoint fields is a stronger requirement that the absence of moduli, but it also generically satisfied.} This is quite an attractive feature for constructing viable models of particle physics and, in particular, semi-realistic models yielding asymptotic freedom \cite{bcms}. On a different spirit, one can think of using these D7-branes to build confining hidden sectors which, via gaugino condensation, generate a non-perturbative superpotential for K\"ahler moduli \cite{gktt}.\footnote{Notice that an Euclidean D3-brane wrapping such moduli-free four-cycle would also contribute  to a non-perturbative superpotential for K\"ahler moduli.}

Eventually, one may conceive building flux compactifications with semi-realistic chiral sectors arising from open strings stretched between D7-branes. A simple example of such construction was given in \cite{ms}, where an MSSM-like model was constructed by using three sets of D7-branes, wrapped on four-cycles $\cs_4^a$, $\cs_4^b$ and $\cs_4^c$ of a $\T^6/(\ZZ_2 \times \ZZ_2)$ orientifold background. The gauge group arising from such D7-branes is the $SU(4) \times SU(2) \times SU(2)$ Pati-Salam extension of the Standard Model, and a simple choice of gauge bundle on the $SU(4)$ D7-brane provides the desired $\cn=1$ chiral spectrum with 3 fermion generations and a minimal Higgs sector. See figure \ref{model} for a schematic representation of this model.

\EPSFIGURE{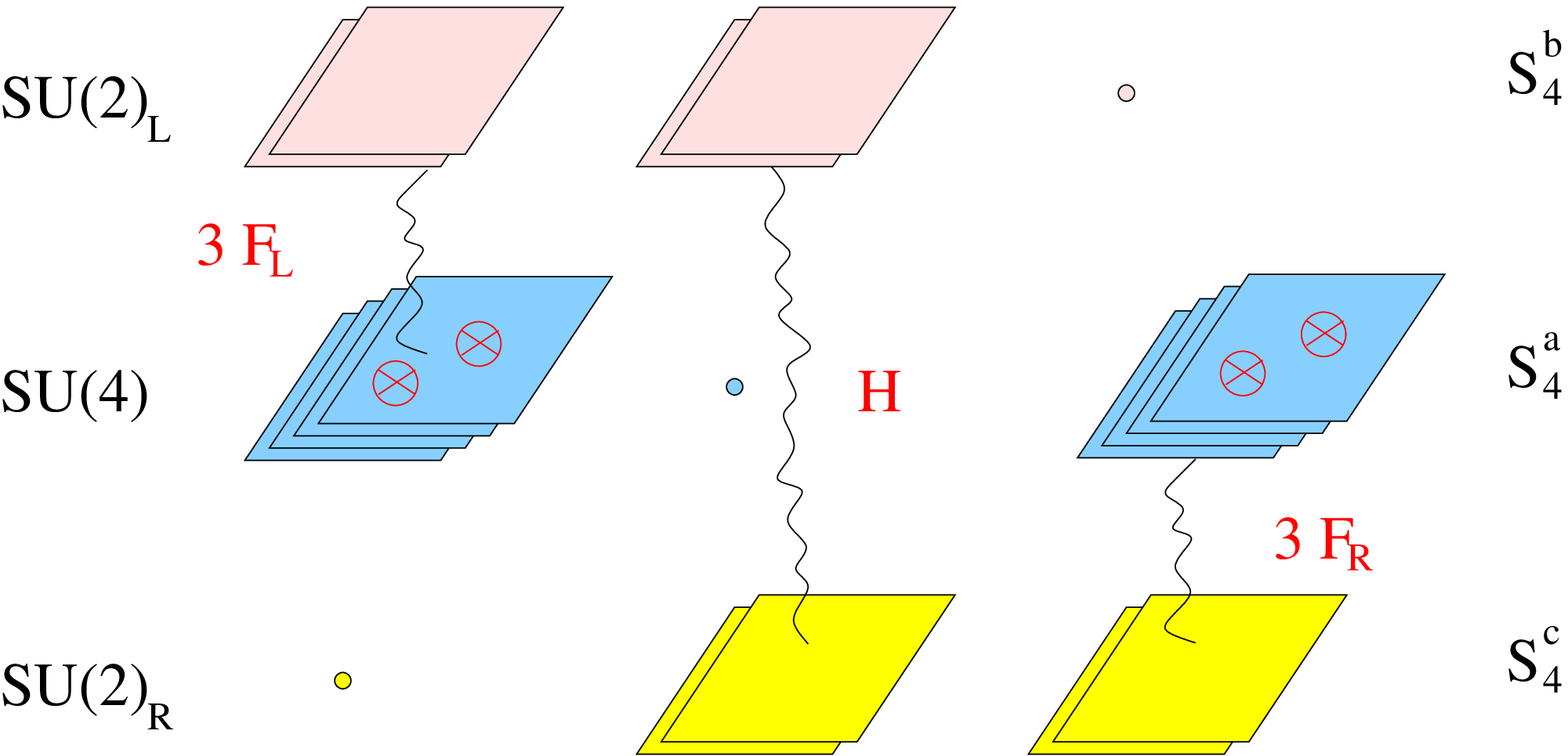, width=4.5in}
{\label{model} Pati-Salam D7-brane model. The open strings with both ends on the same D-brane yield the gauge group $SU(4) \times SU(2)_L \times SU(2)_R$, while those ending on different yield the bifundamental matter $F_L = (4,2,1)$, $F_L = (\bar{4},1,2)$ and $H = (1,2,2)$. Chirality and family replication are due to a non-trivial worldvolume flux $F=dA$ on the $SU(4)$ D7-brane.}

This construction is remarkably simple and, in principle, one can extend it to more general metric backgrounds by choosing three four-cycles $\cs_4^a$, $\cs_4^b$ and $\cs_4^c$ with triple intersection number $\cs_4^a \cap \cs_4^b \cap \cs_4^c = 1$ and appropriate gauge bundles. However, in order to discuss realistic particle physics one needs to connect this Pati-Salam theory with the Standard Model. The usual approach is to perform the adjoint Higgsing $SU(4) \raw SU(3) \times U(1)$ which, in compactifications without background $G_3$ fluxes, is a flat direction of the theory. Now, the field $\Phi$ which acquires a vev is nothing but a massless adjoint field of the theory, so this mechanism is not entirely satisfactory from a phenomenological point of view.

The existence of a D7-brane discretum, however, adds new possibilities for adjoint Higgsing. If our $SU(4)$ D7-brane is placed in a ISD flux background, one can realize the Pati-Salam breaking $SU(4) \raw SU(3) \times U(1)$ by `discrete' adjoint Higgsing, just by first considering 4 D7-branes located on top of each other and then placing one of them in a different supersymmetric location. Notice that the former flat direction is a geometric modulus which is now lifted by the presence of the flux, so $\Phi$ is now given by a massive adjoint field whose vev is at the minimum of a scalar potential. This indeed mimics the field theory mechanism for Pati-Salam adjoint breaking.

\EPSFIGURE{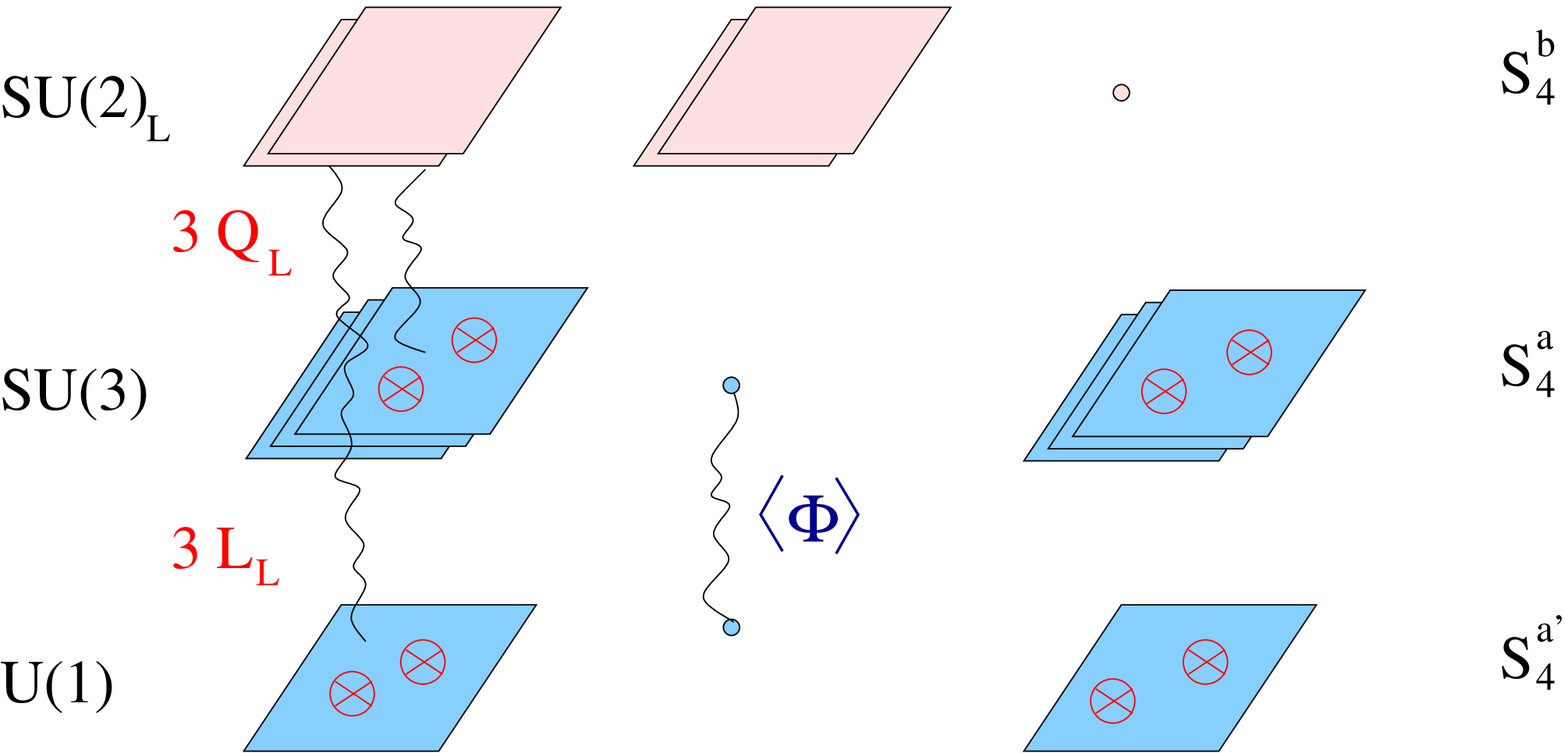, width=4.5in}
{\label{higgs} Pati-Salam breaking $SU(4) \raw SU(3) \times U(1)$
by `discrete' adjoint Higgsing.}

\vspace*{2cm}

\bigskip

\centerline{\bf Acknowledgments}

\bigskip

We wish to thank Sujay Ashok, Pablo G. C\'amara, Marco Gualtieri, Manfred Herbst, Luis E. Ib\'a\~nez, Marcos Mari\~no, Luca Martucci, Rob Myers, Ra\'ul Rabad\'an, Gary Shiu, Joan Sim\'on, and specially \'Angel Uranga for useful comments and discussions. The work of F.M was supported in part by NSF CAREER Award No. PHY-0348093, DOE grant DE-FG-02-95ER40896, and a Research Innovation Award from Research Corporation. FM would also like to thank the Perimeter Institute for Theoretical Physics for hospitality while part of this work was done.

\newpage

\section{D7-brane energetics}
\label{lagrangian}

A non-trivial check of the supersymmetry conditions derived from $\k$-symmetry can be performed by considering the Lagrangian density associated to a D7-brane. Indeed, if we are moving along the moduli space of a supersymmetric D-brane, its Lagrangian density (i.e., its action) should remain constant. On the other hand, if we deform our D-brane in such a way that we break supersymmetry, in general we would expect that the total energy of such D-brane increases after the deformation.

Let us see how this works for a D7-brane in presence of a flux background. The bosonic part of a D7-brane effective action is given by
\be
S_{D7}\, =\, - \int_{M_4 \times \cs_4} g_s^{-1} \sqrt{|g +
\cf|} - \int_{M_4 \times \cs_4} \sum_n C_{2n} \wedge e^\cf
\label{action}
\ee
where $g$ is the induced metric on the D7-brane worldvolume, which is wrapping the four-cycle $\cs_4$, and $\sum_n C_{2n}$ is the usual sum of $C_{2n}$ RR potentials, $n = 0,\dots,4$, which are completed with powers of $\cf$ up to an 8-form.

As we have seen in subsection \ref{kappa}, the fact that a D-brane satisfies the $\kappa$-symmetry conditions allow us to express its contribution to the DBI part of the action by a simple set of $p$-forms evaluated over the D-brane worldvolume. In particular, for the case of a D7-brane, wrapping a holomorphic four-cycle $\cs_4$ and having an anti-selfdual (ASD) field strength $\cf$ allows to write (\ref{action}) as
\bea \label{susyaction}
S_{D7}& = & - \int_{M_4 \times \cs_4} \Phi_{DBI}\, -\,  \int_{M_4 \times \cs_4} \Phi_{WZ} \\ \nonumber 
& = & - \int_{M_4 \times \cs_4} g_s^{-1} \D^{-2} d{\rm vol}_{M_4} \wedge \oh \left(J^2 - \cf^2 \right) \, -\, \int_{M_4 \times \cs_4} \sum_n C_{2n} \wedge e^\cf
\eea
where $\D^{-2} d{\rm vol}_{M_4} = \D^{-2} dx^0 \wedge dx^1 \wedge dx^2 \wedge dx^3$ is the contribution to the DBI energy from the non-compact dimensions wrapped by the D7-brane. It is clear that if we deform our D7-brane embedding as $\cs_4 \raw \cs_4'$, such that $\cs_4'$ is still holomorphic and $\cf$ is an ASD two-form in $\cs_4'$, the expression (\ref{susyaction}) will not change, and now we should evaluate $\Phi_{DBI} + \Phi_{WZ}$ on the worldvolume $M_4 \times \cs_4'$ in order to compute $S_{D7}$. Since $\cs_4$ and $\cs_4'$ both belong to the same homology class $[\cs_4] \in H_4(\cs_4,\ZZ)$, the only thing that we need to check in order to show that $S_{D7}$ does not change is that $\Phi_{DBI} +\Phi_{WZ}$ is a closed form. That is,
\be
d \left(\Phi_{DBI} + \Phi_{WZ}\right)\, = \, 0
\label{cali}
\ee
so that $S_{D7}$ will not change when we move inside $[\cs_4]$ and, at the same time, we respect the $\k$-symmetry conditions.

Let us see that this is indeed the case. Let us first recall that type IIB flux compactifications with warped ansatz (\ref{warped0}), $G_{n\ov{m}} = \D(Z) G_{n\ov{m}}^{CY}$ and O3$^-$-planes require that
\begin{itemize}
\item[i)] We turn on a 5-form field strength $\tilde{F}_5 = (1 + *_{10})\, d\a(y)\wedge dx^0 \wedge dx^1 \wedge dx^2 \wedge dx^3$, related to the warp factor by $d\a = d(\D^{-2} g_s^{-1})$.
\item[ii)] $G_3$ is an imaginary self-dual (ISD) 3-form of $\cm_6$, i.e., $*_6 G_3 = i G_3$.
\end{itemize}
For simplicity, we will assume a background constant axion-dilaton $\tau = C_0 + i/g_s$. The Wess-Zumino contribution to the D7-brane action and its differential are given by
\bea
\Phi_{WZ} & = & C_8 + C_6 \wedge \cf + \oh C_4 \wedge \cf^2 + \frac{1}{3!}  C_2 \wedge \cf^3 + \frac{1}{4!}  C_2 \wedge \cf^4
\label{WZ} \\
d\Phi_{WZ} & = & \tilde{F}_9 + \tilde{F}_7 \wedge \cf + \oh
\tilde{F}_5 \wedge \cf^2 + \frac{1}{3!}  \tilde{F}_3 \wedge \cf^3
+ \frac{1}{4!}  \tilde{F}_1 \wedge \cf^4
\label{dWZ}
\eea
where
\be
\tilde{F}_{2n+1} = d C_{2n} + H_3 \wedge C_{2n-2}
\label{genfs}
\ee
are the generalized field strengths of type IIB supergravity, related to each other by $\tilde{F}_{2n+1} = *_{10} \tilde{F}_{9-2n}$. Using these duality relations and the fact that $\cf$ has support on the four-cycle $\cs_4$, we arrive at
\bea \nonumber
d\Phi_{WZ}  & = & *_{10} \tilde{F}_3 \wedge \cf + \oh d{\rm vol}_{M_4} \wedge d\a \wedge \cf^2 \\
& = & \oh d{\rm vol}_{M_4} \wedge d \left(\D^{-2} g_s^{-1} \cf^2
\right)
\label{fWZ}
\eea
where in the last equality we have used that the supergravity background satisfies $*_6 G_3 = i G_3$ and $d\a = d (\D^{-2} g_s^{-1})$.

It is now easy to compute the differential of $\Phi_{DBI} + \Phi_{WZ}$
\bea
d \left(\Phi_{DBI} + \Phi_{WZ} \right) & = &
\oh {\rm vol}_{M_4} \wedge d \left[g_s^{-1} \D^{-2} (J^2 - \cf^2 + \cf^2)\right] \nonumber \\
& = & \oh g_s^{-1} {\rm vol}_{M_4} \wedge d (J^{\mt{CY}})^2
\label{closed}
\eea
where we have used the explicit expression of the two-form $J = \D\, J^{\mt{CY}}$ in terms of the unwarped Calabi-Yau K\"ahler form $J^{\mt{CY}}$, and the fact that $g_s$ is constant.\footnote{Notice that, had we not assumed a constant axion-dilaton $\tau = C_0 + i/g_s$, then we should also have introduced new sources on the Wess-Zumino part of the action $\int \Phi_{WZ}$.} Clearly (\ref{closed}) vanishes, because $J^{\mt{CY}}$ is a closed form. Hence we obtain that $S_{D7}$ does not change if the $\k$-symmetry conditions are satisfied.

On the other hand, when we break the supersymmetry conditions we would expect the D7-brane action to increase in value. Indeed, notice that whenever $\cf$ is not anti-selfdual $J^2 - \cf^2$ does no longer compute the tension associated to the four-cycle $\cs_4$. In particular, if we consider a self-dual $\cf = *_4 \cf$, such tension would be given by the integral of
\be
\sqrt{|g^{\cs_4} + \cf|} 
= \oh \left. \left(J^2 + \cf \wedge *_4 \cf\right)\right|_{\cs_4} =
\oh \left.\left(J^2 - \cf^2\right)\right|_{\cs_4} 
+ \cf \wedge *_4 \cf \geq \oh \left.\left(J^2 - \cf^2\right)\right|_{\cs_4}
\ee
with the equality being saturated only for $\cf = 0$.\footnote{In general, we can prove that the inequality $\sqrt{|g^{\cs_4} + \cf|} \geq \oh \left.(J^2 - \cf^2)\right|_{\cs_4}$  is only saturated by $\cs_4$ being holomorphic and $\cf$ anti-selfdual.} Let us then consider a D7-brane undergoing the deformation $\cs_4 \raw \cs_4'$, with vanishing $\cf$ at $\cs_4$ and a self-dual two-form $\cf$ at $\cs_4'$. By performing similar computations as above, it is easy to see that the change in the bosonic action is given by
\bea \nonumber
S_{D7} - S_{D7}' & = &
\int_{M_4 \times \cs_4'} \Phi_{DBI} + \Phi_{WZ} + g_s^{-1} \D^{-2} d{\rm vol}_{M_4} \wedge \cf^2
- \int_{M_4 \times \cs_4} \Phi_{DBI} + \Phi_{WZ}\\
 & = &  \int_{\cs_4'} g_s^{-1} \D^{-2} \cf^2
\eea
where in the second line we have normalized the action with respect to the non-compact dimensions, and we have used the fact that $\Phi_{DBI} + \Phi_{WZ}$ is a closed form.

Notice that the computations above remind of those in \cite{gencali}, where it was performed an analysis of supersymmetric D-branes by means of generalized calibrations. Indeed, in the notation of \cite{gencali}, $\Phi_{DBI} +\Phi_{WZ}$ would be the generalized calibration for D7-branes on a compactification with fluxes, and the supersymmetric D7-branes would be the ones that are calibrated with respect to it. As we know, the fact that $\Phi_{DBI} +\Phi_{WZ}$ is a calibration not only implies that it is closed, but also that any D-brane which is not calibrated by it will have higher energy than the calibrated ones. In the present case, D7-branes wrapping holomorphic four-cycles and with an anti-selfdual two-form $\cf$ are the ones that minimize their energy, as we would expect from the fact that they are supersymmetric.

It is important to see that, from the point of view of $\k$-symmetry, the supersymmetry conditions for a type IIB D-brane are given in terms of the non-closed forms $\Om$ and $J$ defined on a general $SU(3)$-structure manifold, rather than by a generalized calibration. As advanced in \cite{gencali} and checked here, both approaches are not in contradiction, but rather are complementary approaches to characterize BPS D-branes in $SU(3)$-structure compactifications. In fact, the $\k$-symmetry equations derived in subsection \ref{kappa} are a non-trivial refinement of the proposal made in \cite{gencali}.

\end{document}